\documentclass[conference]{IEEEtran}
\IEEEoverridecommandlockouts
\usepackage{cite}
\usepackage{amsmath,amssymb,amsfonts}
\usepackage{algorithmic}
\usepackage{graphicx}
\usepackage{textcomp}
\usepackage{xcolor}

\usepackage{hyperref}
\usepackage{array}
\usepackage{booktabs}
\usepackage{url}

\def\BibTeX{{\rm B\kern-.05em{\sc i\kern-.025em b}\kern-.08em
    T\kern-.1667em\lower.7ex\hbox{E}\kern-.125emX}}
\begin{document}

\title{
A Regressor-Guided Graph Diffusion Model for Predicting Enzyme Mutations to Enhance Turnover Number
}

\author{\IEEEauthorblockN{Xiaozhu Yu\textsuperscript{1,2,$\S$}, Kai Yi\textsuperscript{1,3}, Yu Guang Wang\textsuperscript{1,4}, Yiqing Shen\textsuperscript{1,5,*}}
\IEEEauthorblockA{
\textsuperscript{1}\textit{Toursun Synbio}, Shanghai, China\\
\textsuperscript{2}\textit{Pratt School of Engineering}, \textit{Duke University}, Durham, USA\\
\textsuperscript{3}\textit{MRC Laboratory of Molecular Biology}, \textit{University of Cambridge}, Cambridge, UK\\
\textsuperscript{4}\textit{Institute of Natural Sciences}, \textit{Shanghai Jiao Tong University}, Shanghai, China\\
\textsuperscript{5}\textit{Department of Computer Science}, \textit{Johns Hopkins University}, Baltimore, USA\\
{\footnotesize \textsuperscript{$\S$}Work done during the internship at Toursun Synbio. 
\textsuperscript{*}Corresponding Author.}\\
{\footnotesize xiaozhu.yu@duke.edu\qquad yiqingshen1@gmail.com}}
}

\maketitle

\begin{abstract}
Enzymes are biological catalysts that can accelerate chemical reactions compared to uncatalyzed reactions in aqueous environments. 
Their catalytic efficiency is quantified by the turnover number ($k_{cat}$), a parameter in enzyme kinetics.
Enhancing enzyme activity is important for optimizing slow chemical reactions, with far-reaching implications for both research and industrial applications.
However, traditional wet-lab methods for measuring and optimizing enzyme activity are often resource-intensive and time-consuming.
To address these limitations, we introduce $k_{cat}$Diffuser, a novel regressor-guided diffusion model designed to predict and improve enzyme turnover numbers. 
Our approach innovatively reformulates enzyme mutation prediction as a protein inverse folding task, thereby establishing a direct link between structural prediction and functional optimization. 
$k_{cat}$Diffuser is a graph diffusion model guided by a regressor, enabling the prediction of amino acid mutations at multiple random positions simultaneously.
Evaluations on BERENDA dataset shows that $k_{cat}$Diffuser can achieve a $\Delta \log k_{cat}$ of 0.209, outperforming state-of-the-art methods like ProteinMPNN, PiFold, GraDe-IF in improving enzyme turnover numbers.
Additionally, $k_{cat}$Diffuser maintains high structural fidelity with a recovery rate of 0.716, pLDDT score of 92.515, RMSD of 3.764, and TM-score of 0.934, demonstrating its ability to generate enzyme variants with enhanced activity while preserving essential structural properties.
Overall, $k_{cat}$Diffuser represents a more efficient and targeted approach to enhancing enzyme activity. 
The code is available at \url{https://github.com/xz32yu/KcatDiffuser}.
\end{abstract}


\section{Introduction}
Enzymes are biological catalysts that accelerate chemical reactions and maintaining cellular metabolic processes essential for life \cite{b1}. 
These protein molecules lower the activation energy of biochemical reactions, enabling them to occur at rates compatible with cellular function.
The efficiency of enzyme catalysis is often quantified by the turnover number ($k_{cat}$), a key parameter in enzyme kinetics that provides insights into cellular metabolism, proteome allocation, and physiological diversity \cite{b2}. 
$k_{cat}$ represents the maximum number of substrate molecules converted to product per enzyme molecule per unit time under saturating substrate conditions.
However, experimental determination of $k_{cat}$ is both time-consuming and resource-intensive, requiring purified enzymes and specialized equipment \cite{b3}. 
This limitation has led to a scarcity of experimentally measured $k_{cat}$ values, with less than 1\% of enzymes listed in the UniProt database having experimentally determined $k_{cat}$ values \cite{b5}.

Recent advancements in artificial intelligence, particularly deep learning, have led to the emergence of models capable of predicting enzyme activity from various inputs.
For instance, DLKcat can predict metabolic enzyme activity based on substrate structure and enzyme sequence \cite{b4}. 
This model utilizes a combination of convolutional neural networks and graph neural networks to process protein sequences and substrate structures, respectively.
Building upon this work, DeepEnzyme incorporates enzyme protein structure as an additional input to enhance prediction accuracy \cite{b5}.
By leveraging the integrated features from both sequences and 3D structures, DeepEnzyme demonstrates improved robustness when processing enzymes with low sequence similarity compared to those in the training dataset \cite{b5}.
While these models have shown promise in predicting $k_{cat}$ values, they primarily focus on estimating existing enzyme activities rather than addressing the crucial challenge of improving enzyme activity.

In the field of protein design, models such as Evolutionary Scale Modeling-1v (ESM-1v) have been developed to predict the effects of protein variants  on a wide range of properties \cite{b6}. 
These models leverage large-scale protein sequence data to learn evolutionary patterns and make zero-shot predictions of mutational effects across diverse proteins with different functions \cite{b6}.
While ESM-1v and similar models have shown promise in predicting variant effects, they are not specifically designed to optimize enzyme kinetic parameters like the turnover number ($k_{cat}$) or suggest mutations for enhancing enzyme activity.
Traditional methods that rely on single or double amino acid substitutions often fail to achieve significant improvements in enzyme activity due to the complex interdependencies within protein structures.
To address the limitations of current approaches, we focus on improving enzyme activity in this paper by proposing a model capable of modifying amino acids at multiple random positions, a task well-suited for diffusion models \cite{b7}.
Diffusion models have shown promise in the field of protein design. Notably, GraDe-IF \cite{b7} has emerged as a powerful model for inverse protein folding, which aims to maintain the given protein backbone while generating new amino acid sequences.
GraDe-IF's ability to produce diverse sequences while preserving structural integrity makes it an ideal starting point for our work on enzyme mutation.
By adapting and extending the principles of GraDe-IF, we aim to develop a diffusion-based model specifically tailored to enhance enzyme turnover numbers.

The major contributions of this work is three fold. 
Firstly, we innovatively reformulate enzyme mutation prediction for optimizing turnover number as a protein inverse folding task, thereby establishing a direct link between structural prediction and functional optimization.
Secondly, we introduce a regressor-guided graph diffusion model, named $k_{cat}$Diffuser, designed to enhance turnover number ($k_{cat}$). 
Moreover, we intergate it with an efficient sampling scheme using DDIM, allowing for larger step sizes and faster generation of enzyme variants. 
$k_{cat}$Diffuser enables modifications of amino acids at multiple random positions simultaneously, overcoming the limitations of traditional site-directed mutation prediction methods.
Finally, we train our model on the BRENDA enzyme dataset, ensuring its applicability to a wide range of enzymatic systems and demonstrating its potential for generalizable enzyme optimization.

\begin{figure}[htbp]
    \centering
    \includegraphics[width=\linewidth]{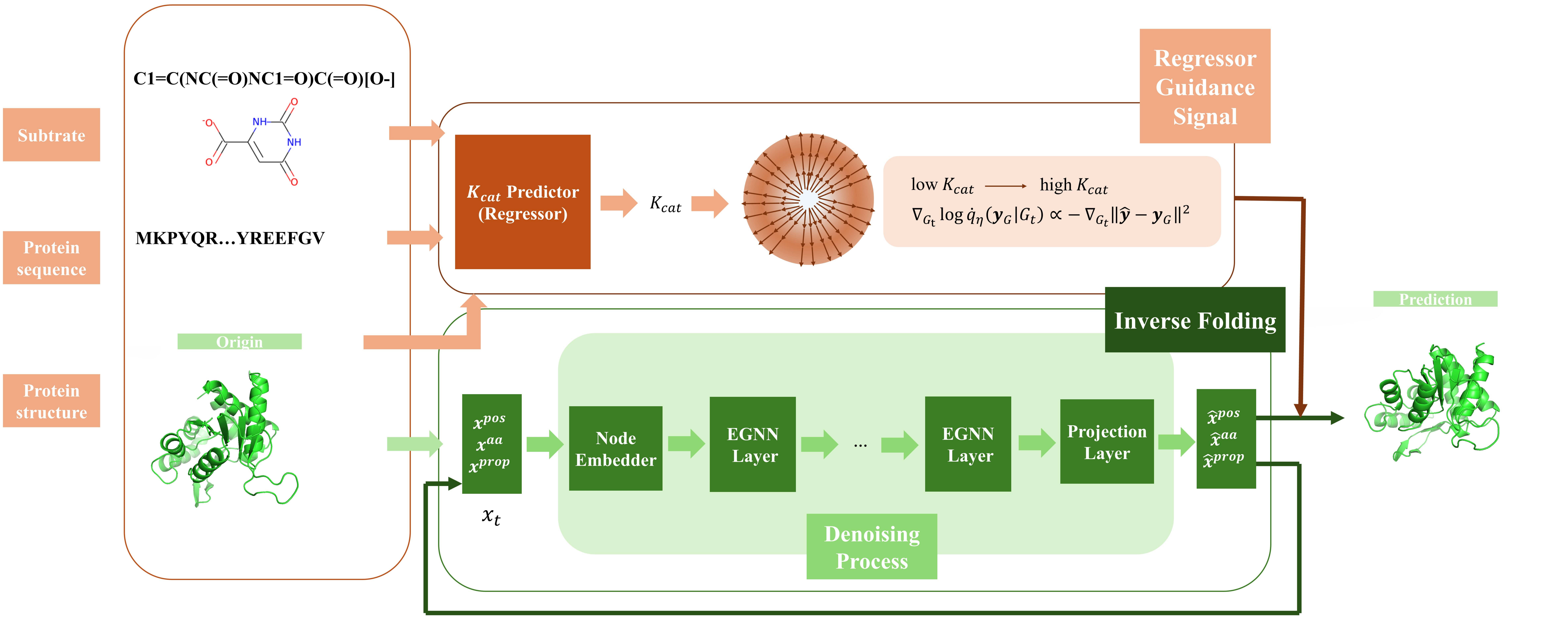}
    \caption{Overview of $k_{cat}$Diffuser. The framework combines an inverse protein folding diffusion model with a $k_{cat}$ regressor for guided sampling. 
    The input consists of a substrate, protein sequence, and protein structure. 
    The inverse folding component uses a graph-based diffusion model to generate new amino acid sequences. 
    Concurrently, the $k_{cat}$ predictor (regressor) estimates the turnover number, providing a guidance signal to steer the denosing process towards sequences with potentially higher $k_{cat}$ values. 
    The regressor guidance is implemented through a gradient-based approach, pushing the sampling towards regions of higher predicted $k_{cat}$.}\label{fig:framework}
\end{figure}

\section{Methods}

\subsection{Problem Formulation}
This work aims to develop a method for generate enzyme variants that enhance the turnover number ($k_{cat}$). 
We formulate it as an inverse protein folding problem with an additional optimization objective during the sampling stage with a regressor guidance signal. 
Given a protein structure represented by its backbone coordinates $\boldsymbol{X}^{pos} = {x^{pos}_1, \ldots, x^{pos}_n}$, where $n$ is the number of amino acids, our task is to predict a set of feasible amino acid sequences $\boldsymbol{X}^{aa} = {x^{aa}_1, \ldots, x^{aa}_n}$ that can fold into the given structure, while simultaneously identifying sequences likely to exhibit improved $k_{cat}$ values compared to the wild-type enzyme.
We approach this problem by modeling the conditional probability distribution $p(\boldsymbol{X}^{aa} | \boldsymbol{X}^{pos})$, which represents the likelihood of amino acid sequences given the backbone structure. 
To incorporate the optimization of $k_{cat}$, we introduce a regressor function $g_{\eta}(\boldsymbol{X}^{aa}, \boldsymbol{X}^{pos}) \rightarrow k_{cat}$ that predicts the turnover number for a given sequence and structure.
To solve it, we develop $k_{cat}$Diffuser, a regressor-guided graph diffusion model, as depicted in Fig.~\ref{fig:framework}. 
This model combines a protein inverse folding diffusion model $p_{\theta}(\boldsymbol{X}^{aa} | \boldsymbol{X}^{pos})$ that generates diverse amino acid sequences compatible with the given backbone structure, and a regressor $g_{\eta}$ that guides the sampling process towards sequences with potentially higher $k_{cat}$ values.
In our framework, we represent the protein structure as a graph $G = \{\boldsymbol{X}, \boldsymbol{A}, \boldsymbol{E}\}$, where $\boldsymbol{X}$ includes positional, amino acid type, and physicochemical property information, $\boldsymbol{A}$ is the adjacency matrix, and $\boldsymbol{E}$ captures edge features. 
This graph representation allows us to capture the complex spatial relationships within the protein structure.
By integrating these components, $k_{cat}$Diffuser aims to generate enzyme variants that not only maintain the desired protein structure but also exhibit enhanced catalytic efficiency as measured by $k_{cat}$.

\subsection{Protein Graph Construction and Feature Encoding}
To implement $k_{cat}$Diffuser, we represent proteins as graphs $G = \{\boldsymbol{X}, \boldsymbol{A}, \boldsymbol{E}\}$ by converting PDB files into this format \cite{b7}. 
The node features $\boldsymbol{X}$ are defined as:
\begin{equation}
\boldsymbol{X} = [\boldsymbol{X}^{pos}, \boldsymbol{X}^{aa}, \boldsymbol{X}^{prop}],
\end{equation}
where $\boldsymbol{X}^{pos}$ denotes the 3D coordinates of the $\alpha$-carbon, $\boldsymbol{X}^{aa}$ is a one-hot encoded vector of amino acid type, and $\boldsymbol{X}^{prop}$ represents physicochemical properties.
Edge attributes $\boldsymbol{E}$ capture spatial and chemical relationships between connected residues, \textit{i}.\textit{e}.
\begin{equation}
\boldsymbol{E} = [d_{ij}, \Delta pos_{ij}, \phi_{ij}],
\end{equation}
where $d_{ij}$ is the distance between residues $i$ and $j$, $\Delta pos_{ij}$ is their relative position, and $\phi_{ij}$ encodes dihedral angles.
The graph construction utilizes a k-nearest neighbor (kNN) method, establishing connections between amino acids within a 30\AA radius to preserve the protein's tertiary structure while creating a computationally tractable representation.
The corresponding adjacency matrix $\boldsymbol{A}$ is defined by
\begin{equation}
A_{ij} =
\begin{cases}
1 & \text{if } |\boldsymbol{X}^{pos}_i - \boldsymbol{X}^{pos}_j| < 30\text{\AA} \text{ and } j \in \text{kNN}(i) \\
0 & \text{otherwise}
\end{cases}.
\end{equation}
For better graph representation, we further incorporate protein backbone information, including dihedral angles ($\psi$, $\phi$) and secondary structure elements (ss), encoded as:
\begin{equation}
\boldsymbol{X}_{backbone} = [\cos(\psi), \sin(\psi), \cos(\phi), \sin(\phi), \text{one\_hot}(\text{ss})].
\end{equation}

\subsection{Protein Inverse Folding Diffusion Model}
In the protein inverse folding diffusion model $p_{\theta}$, the diffusion process gradually adds noise to the amino acid types $\boldsymbol{X}^{aa}$ over $T$ timesteps, transforming them from the original sequence to a uniform distribution. 
This process is defined by a forward transition probability $q(\boldsymbol{x}_t | \boldsymbol{x}_{t-1})$, where $\boldsymbol{x}_t$ represents the noisy amino acid types at timestep $t$. 
The reverse denoising process, parameterized by $\theta$, aims to recover the original sequence $\boldsymbol{X}^{aa}$ through iterative refinement
\begin{equation}
p_{\theta}(\boldsymbol{x}_{t-1} | \boldsymbol{x}_t) \propto \sum_{\boldsymbol{x}^{aa}} q(\boldsymbol{x}_{t-1} | \boldsymbol{x}_t, \boldsymbol{x}^{aa}) 
\cdot p_{\theta}(\boldsymbol{x}^{aa} | \boldsymbol{x}_t),
\end{equation}
where, $p_{\theta}(\boldsymbol{x}^{aa} | \boldsymbol{x}_t)$ is predicted by an equivalent graph neural network (EGNN) as the denoising network that takes as input the noisy protein graph and additional structural information including backbone dihedral angles and secondary structure elements.
To accelerate sampling, we employ the Denoising Diffusion Implicit Models (DDIM) \cite{b34}, which allows for larger step sizes in the sampling process:
\begin{equation}
p_{\theta}(\boldsymbol{x}_{t-k}|\boldsymbol{x}_t) \propto (\sum_{\boldsymbol{x}^{aa}} q(\boldsymbol{x}_{t-k}|\boldsymbol{x}_t,\boldsymbol{x}^{aa})\hat{p}(\boldsymbol{x}^{aa}|\boldsymbol{x}_t))^T
\end{equation}
where $T$ controls the time step. 
The multi-step posterior $q(\boldsymbol{x}_{t-k}|\boldsymbol{x}_t,\boldsymbol{x}^{aa})$ is computed using the cumulative transition matrices
\begin{equation}
\begin{aligned}
q(\boldsymbol{x}_{t-k}|&\boldsymbol{x}_t,\boldsymbol{x}^{aa})  = \\
&\text{Cat}\Big(\boldsymbol{x}_{t-k}|\frac{\boldsymbol{x}_t Q^T_t...Q^T_{t-k} \odot \boldsymbol{x}^{aa}\bar{Q}_{t-k}}{\boldsymbol{x}^{aa}\bar{Q}_t \boldsymbol{x}^T_t}\Big),
\end{aligned}
\end{equation}
where $Q_t$ represents the transition matrix at time $t$, and $\bar{Q}_t$ is the cumulative transition matrix up to time $t$ 
In this equation, $\text{Cat}(\cdot)$ denotes the categorical distribution, while $\odot$ represents element-wise multiplication (Hadamard product).
The fraction inside $\text{Cat}(\cdot)$ represents the unnormalized probabilities for the categorical distribution, with $Q^T_t$ being the transpose of the transition matrix at time $t$.

\subsection{Regressor-Guided Graph Diffusion Sampling for $k_{cat}$ Optimization}
We introduce a regressor-guided diffusion sampling scheme that combines a $k_{cat}$ regressor with the protein inverse folding diffusion model to efficiently generate enzyme variants with potentially improved $k_{cat}$ values.

\paragraph{Regressor}
The regressor $g_{\eta}$ comprises of Transformer \cite{b32} and GCN \cite{b33} to extract features from both protein 1D sequences and 3D structures.
It processes overlapping n-grams of amino acids through an embedding layer and a Transformer encoder, while the GCN processes the 3D structure inputs of the protein graph and substrate. 
The regressor $g_{\eta}$ then fuses these representations to predict the $k_{cat}$ value. 
We train the learnable weights $\eta$ on the BRENDA dataset of enzyme kinetic data.

\begin{figure*}[htbp]
    \centering
    \includegraphics[width=0.75\textwidth]{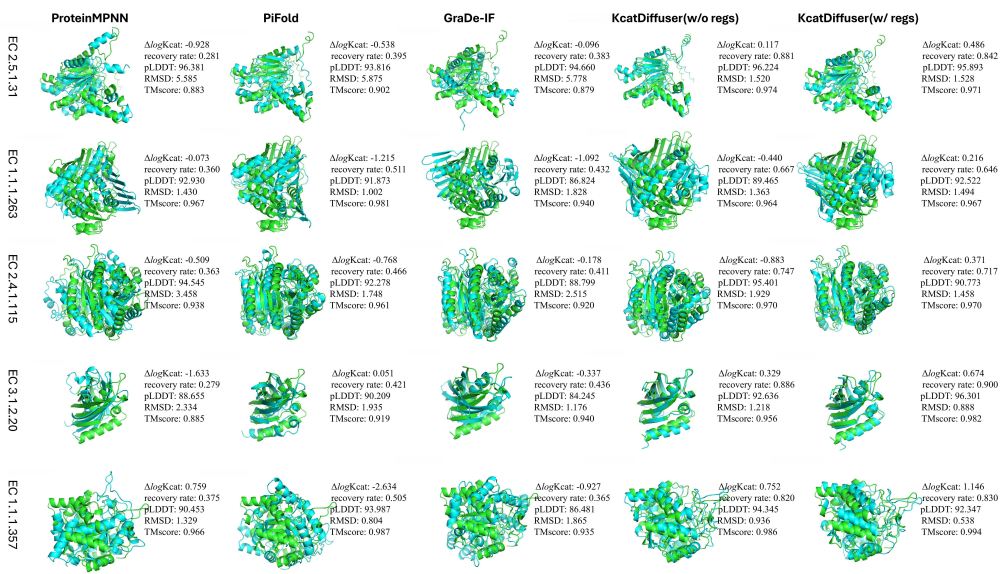}
    \caption{Case study comparison of protein generated by different models. Each row represents a distinct enzyme (EC numbers shown on the left). 
    Columns show results from ProteinMPNN, PiFold, GraDe-IF, and $k_{cat}$Diffuser (without and with regressor guidance). 
    Green structures represent the original proteins, while cyan structures are the generated variants. 
    Performance metrics are provided for each case, including $\Delta\log k_{cat}$, recovery rate, pLDDT, RMSD, and TM-score.
    %
    }\label{fig:casestudy}
\end{figure*}

\paragraph{Regressor-guided Sampling}
We use the regressor $g_{\eta}$ to guide the unconditional protein inverse folding diffusion model $\phi_{\theta}$.
The target prediction $\boldsymbol{y}_G$ of a protein graph $G$ is obtained from a noisy version of $G$ where $g_{\eta}(G_t)=\hat{\boldsymbol{y}}$.
We make an assumption that the conditional probability of the noisy sequence given the target property is equal to the unconditional probability. 
This simplification allows us to factorize the joint probability and incorporate the regressor's guidance more effectively. 

Under this assumption, we have $\dot{q}(\boldsymbol{x}_{t-k}|\boldsymbol{x}_t,\boldsymbol{x}^{aa},\boldsymbol{y}_G)=\dot{q}(\boldsymbol{x}_{t-k}|\boldsymbol{x}_t,\boldsymbol{x}^{aa})$, which yields to
\begin{equation}
\dot{q}(\boldsymbol{x}_{t-k}|\boldsymbol{x}_t,\boldsymbol{x}^{aa},\boldsymbol{y}_G)\propto{q(\boldsymbol{x}_{t-1}|\boldsymbol{x}_t,\boldsymbol{x}^{aa})\dot{q}(\boldsymbol{y}_G|\boldsymbol{x}_{t-k})},
\end{equation}
where $\boldsymbol{y}_G$ indicates the target properties of a protein graph $G$, $\dot{q}$ denotes the noising process conditioned on $\boldsymbol{y}_G$, and $q$ denotes the unconditional noising process.
This equation expresses the probability of a less noisy sequence $\boldsymbol{x}_{t-k}$ given the current noisy sequence $\boldsymbol{x}_t$, the original sequence $\boldsymbol{x}^{aa}$, and the target property $\boldsymbol{y}_G$. 
It combines the unconditional noising process $q$ with the conditional probability of the target property given the less noisy sequence.
To make this formulation computationally tractable, we employ a first-order approximation to define $\nabla_{\boldsymbol{x}}$, which allows us to linearize the log-probability around the current point, described as
\begin{equation}
\nonumber
\begin{split}
&\log\dot{q}(\boldsymbol{y}_G|\boldsymbol{x}_{t-k},\boldsymbol{x}^{aa}) \\
&\approx \log\dot{q}(\boldsymbol{y}_G|\boldsymbol{x}_t,\boldsymbol{x}^{aa})+\\
& \qquad \lambda \langle\nabla{\boldsymbol{x}} \log\dot{q}(\boldsymbol{y}_G|\boldsymbol{x}_{t-k},\boldsymbol{x}^{aa}),\boldsymbol{x}_{t-k}-\boldsymbol{x}_t\rangle \\
&\approx c(\boldsymbol{x}_t,\boldsymbol{x}^{aa})+\lambda \sum_{1\leq i\leq n} \langle\nabla{x_i} \log\dot{q}(\boldsymbol{y}_G|\boldsymbol{x}_{t-k},\boldsymbol{x}^{aa}),\boldsymbol{x}_{i,t-1}\rangle
\end{split}
\end{equation}
where $\lambda$ indicates the extent to which the regressor $\boldsymbol{y}_G$ influences the outcomes, and $c$ is a function independent of $\boldsymbol{x}_{t-k}$.
%
Finally, assuming that the conditional probability of the target property follows a Gaussian distribution centered at the regressor's prediction, i.e. $\dot{q}(\boldsymbol{y}_G|\boldsymbol{x}_t,\boldsymbol{x}^{aa})=\mathcal{N}(g(\boldsymbol{x}_t),\sigma_y\boldsymbol{I})$, we can express the gradient of the log-probability with respect to the protein graph as:
\begin{equation}
\nabla_{G_t}\dot{q}_{\eta}(\boldsymbol{y}|G_t)\propto{-\nabla{G_t}|\hat{\boldsymbol{y}}-\boldsymbol{y}_G|^2}
\end{equation}
where $g$ is estimated by $g_{\eta}$.
This gradient guides the sampling process towards protein sequences that are more likely to exhibit the higher $k_{cat}$ value during the inverse folding process.

\begin{table}[t!]
\centering
\caption{Performance comparison in terms of $\Delta \log k_{cat}$ (improvement in enzyme turnover number), Recovery Rate, pLDDT, RMSD, and TM-score. 
Arrows indicate whether higher ($\uparrow$) or lower ($\downarrow$) values are better. $k_{cat}$Diffuser demonstrates superior performance across most metrics, particularly in enzyme activity improvement ($\Delta \log k_{cat}$) and sequence recovery, while maintaining high structural quality.
Best performance of each metric is marked in \textbf{bold}.
}\label{tab:result}
\resizebox{\linewidth}{!}{
\begin{tabular}{lccccc}
\toprule
\textbf{Methods} & \boldmath{$\Delta \log k_{cat}$} $(\uparrow)$ & \textbf{Recovery Rate} $(\uparrow)$ & \textbf{pLDDT} $(\uparrow)$ & \textbf{RMSD} $(\downarrow)$ & \textbf{TM-score} $(\uparrow)$\\
\midrule
\textbf{ProteinMPNN}\cite{b20} & 0.117 & 0.342 & 92.038 & 5.444 & 0.892\\
\textbf{PiFold}\cite{b21}  & 0.087 & 0.473 & \textbf{92.968} & 4.430 & 0.922\\
\textbf{GraDe-IF}\cite{b7} & -0.057 & 0.406 & 89.165 & 7.533 & 0.810\\
\midrule
\textbf{\boldmath{$k_{cat}$Diffuser}} & \textbf{0.209} & \textbf{0.716} & 92.515 & \textbf{3.764} & \textbf{0.934}\\
\bottomrule
\end{tabular}}
\end{table}

\section{Experiments}

\subsection{Implementation Details}
The regressor employs a Transformer-based architecture, comprising 3 output layers and 3 Transformer encoding layers. 
It utilizes an input dimension of 20, a hidden dimension of 64, and 4 attention heads to capture diverse input features effectively.
For the $k_{cat}$Diffuser, we adopted a learning rate of $0.0005$ and a dropout rate of $0.1$ to mitigate overfitting. 
To ensure training stability, gradient clipping was applied with a threshold of $1.0$. The denoising network in $k_{cat}$Diffuser is a EGNN with 6 layers, each with a hidden size of 128 units, and incorporates embedding layers with a dimension of 128.
The training process involves a diffusion sequence length of 500 time steps. 
To enhance the model's robustness and account for sequence variability, we introduced BLOSUM-based noise to the input data \cite{b7}. 
This noise injection simulates natural amino acid substitutions, potentially improving the model's generalization capabilities for enzyme mutation prediction \cite{b7}.

\subsection{Dataset Preparation}
To train $k_{cat}$Diffuser, in addition to the CATH dataset, we also leverage BRENDA enzyme database, which includes EC numbers, organisms, enzyme sequences, simplified molecular-input line-entry system (SMILES) representations of substrates, and $k_{cat}$ values \cite{b26}. 
We focus on enzyme-substrate pairs to align with our model's objective of optimizing $k_{cat}$.
While $k_{cat}$Diffuser requires protein structures as input, BRENDA primarily provides enzyme sequences. 
To bridge this gap, we employ ESMFold \cite{b27} to predict 3D structures from these sequences, resulting in 15,603 enzyme structures. 
We divide this dataset into 12,482 enzymes for training, 1,560 for validation, and 1,561 for testing. 
These structures are then converted into graph representations $G = \{\boldsymbol{X}, \boldsymbol{A}, \boldsymbol{E}\}$ using our pre-processing pipeline.
To ensure data quality and computational feasibility, we filter out empty files and structures larger than 10MB from both CATH and BRENDA datasets before pre-processing. 
We investigate the impact of incorporating the BRENDA dataset by training $k_{cat}$Diffuser on two configurations: CATH dataset only and combined CATH and BRENDA datasets in the experiment section. 
While training configuration differs, we evaluate both configurations on the BRENDA test set. 

\subsection{Evaluation Metrics}
The first evaluation metric is $\Delta \log k_{cat}$, which quantifies the improvement in the enzyme's turnover number. 
A higher value indicates more enhancement in catalytic efficiency. 
Then, we assess the model's ability to generate sequences similar to the native protein using the recovery rate. 
To assess the structural quality of generated sequences, we use ESMFold to predict their 3D structures and compare them to the original crystal structures. 
We evaluate foldability using three metrics: pLDDT, a confidence measure for per-residue structural accuracy \cite{b28}; RMSD (Root Mean Square Deviation), which measures atomic-level differences between model and native structures \cite{b29}; and TM-score (Template Modeling score), which assesses global structural similarity, correlating strongly with overall model quality \cite{b30,b31}. 
These metrics provide a comprehensive evaluation of the generated sequences' ability to maintain the desired protein structure while potentially exhibiting improved $k_{cat}$ values, aligning with the core objectives of $k_{cat}$Diffuser.

\subsection{Results}
The results, summarized in Table \ref{tab:result}, demonstrate the effectiveness of our approach across multiple metrics.
Specifically, $k_{cat}$Diffuser achieved the highest improvement in enzyme turnover number, with a $\Delta \log k_{cat}$ of 0.209. This represents an enhancement over ProteinMPNN (0.117) and PiFold (0.087), while GraDe-IF showed a slight decrease (-0.057), demonstrating the effectiveness of our regressor-guided diffusion approach in optimizing enzyme activity. 
Our $k_{cat}$Diffuser also outperformed all baselines in in terms of recovery rate, achieving 0.716. This is higher than PiFold (0.473), GraDe-IF (0.406), and ProteinMPNN (0.342). 
The high recovery rate indicates that $k_{cat}$Diffuser generates sequences that closely resemble the native protein while still introducing beneficial mutations.
In terms of structural quality, $k_{cat}$Diffuser maintained high fidelity while improving enzyme activity. 
Our model achieved a pLDDT score of 92.515, slightly lower than PiFold (92.968) but higher than ProteinMPNN (92.038) and GraDe-IF (89.165), indicating high confidence in the local structural accuracy of our generated sequences.
$k_{cat}$Diffuser achieved the lowest RMSD of 3.764, better than all baselines, suggesting that our generated structures closely align with the native structures at the atomic level. 
Furthermore, our model attained the highest TM-score of 0.934, indicating excellent global structural similarity to the native proteins.
These results demonstrate that $k_{cat}$Diffuser can balance enzyme activity improvement with structural integrity. 
To further illustrate the performance of $k_{cat}$Diffuser, we conducted a case study across five diverse enzyme classes (Fig.~\ref{fig:casestudy}). 
The visual comparison and accompanying metrics demonstrate the superiority of our approach, particularly when using regressor guidance.
Across all cases, $k_{cat}$Diffuser consistently achieved higher $\Delta \log k_{cat}$ values, indicating greater improvements in enzyme activity. 
For instance, in the EC 2.5.1.31 case, $k_{cat}$Diffuser with regressor guidance achieved a $\Delta \log k_{cat}$ of 0.486, outperforming other methods. 
Importantly, these activity improvements were achieved while maintaining high structural fidelity, as evidenced by the consistently low RMSD values and high TM-scores. 
The generated structures (cyan) closely align with the original proteins (green), demonstrating $k_{cat}$Diffuser's ability to optimize enzyme activity without compromising structural integrity. 

\begin{table}[htbp!]
\centering
\caption{Comparison of model complexity. The table shows the number of parameters in millions, memory usage in megabytes, and inference time in seconds for each compared model.
}\label{tab:complexity}
\resizebox{\linewidth}{!}{%
\begin{tabular}{ccccc}
\toprule
\textbf{Methods} & \# Param. (M) & Memory (MB) & Time (s)\\
\midrule
ProteinMPNN & 1.66 & 237.1 & 0.60 \\
PiFold  & 6.61 & 108.0 & 0.26 \\
GraDe-IF & 7.64 & 140.9 & 0.39 \\
\midrule
$k_{cat}$Diffuser & 8.85 & 170.0 & 5.23 \\
\bottomrule
\end{tabular}
}
\end{table}

\subsection{Model Complexity Analysis}
We conducted a comprehensive analysis of model complexity, comparing $k_{cat}$Diffuser with ProteinMPNN, PiFold, and GraDe-IF. 
Table \ref{tab:complexity} summarizes the results in terms of number of parameters, memory usage, and inference time.
$k_{cat}$Diffuser has the highest number of parameters (8.85M) among the compared models.
This increased complexity allows $k_{cat}$Diffuser to capture relationships between protein structure and enzyme activity.
In terms of memory usage, $k_{cat}$Diffuser (170.0 MB) sits between the memory-efficient PiFold (108.0 MB) and the more memory-intensive ProteinMPNN (237.1 MB). 
This moderate memory footprint makes KcatDiffuser suitable for deployment on a wide range of hardware configurations, balancing performance with resource requirements.
The inference time of $k_{cat}$Diffuser (5.23s) is notably higher than the other models, which range from $0.26$s to $0.60$s. 
This increased computational cost is primarily due to the iterative nature of the diffusion process and the additional computations required for the regressor-guided sampling. 
However, this trade-off in speed enables $k_{cat}$Diffuser to perform multi-site mutations and optimize for enzyme activity, capabilities not present in the faster models.

\begin{table}[t!]
\centering
\caption{Ablation study on the influence of regressor guidance strength ($\lambda$) on $k_{cat}$Diffuser performance.}\label{tab:ablation}
\resizebox{\linewidth}{!}{
\begin{tabular}{lccccc}
\toprule
\textbf{$\lambda$} & \boldmath{$\Delta \log k_{cat}$} ($\uparrow$)& \textbf{Recovery Rate} ($\uparrow$) & \textbf{pLDDT} ($\uparrow$) & \textbf{RMSD} ($\downarrow$) & \textbf{TM-score} ($\uparrow$)\\
\midrule
\textbf{0.1} & 0.194 & 0.730 & 92.343 & 3.641 & 0.937\\
\textbf{0.5} & 0.167 & 0.732 & 92.403 & 3.626 & 0.937\\
\textbf{1.0} & 0.184 & 0.733 & 92.408 & 3.638 & 0.937\\
\textbf{5.0} & 0.209 & 0.716 & 92.515 & 3.764 & 0.934\\
\textbf{10.0} & 0.124 & 0.643 & 91.408 & 6.253 & 0.867\\
\textbf{20.0} & 0.083 & 0.537 & 85.920 & 16.390 & 0.613\\
\bottomrule
\end{tabular}
}
\end{table}

\subsection{Ablation Study}
To explore the impact of regressor guidance on $k_{cat}$Diffuser's performance, we conducted an ablation study by varying the regressor guidance strength parameter $\lambda$. 
Table \ref{tab:ablation} presents the results of this study, demonstrating the trade-off between enzyme activity improvement and structural integrity.
For lower values of $\lambda$ (0.1 to 1.0), we observe relatively stable performance across all metrics. The model maintains high structural fidelity, as evidenced by consistent pLDDT scores around 92.4, low RMSD values (3.64), and high TM-scores (0.937). 
The recovery rates are also highest in this range (0.730-0.733), indicating that the generated sequences closely resemble the native proteins.
As $\lambda$ increases to 5.0, we see an improvement in $\Delta \log k_{cat}$ (0.209), suggesting enhanced enzyme activity. 
This comes with a slight decrease in recovery rate (0.716) and marginal changes in structural metrics, indicating a good balance between activity improvement and structural preservation.
However, further increases in $\lambda$ (10.0 and 20.0) lead to a decline in performance across all metrics. 
The $\Delta \log k_{cat}$ decreases, and we observe a marked deterioration in structural integrity, particularly at $\lambda=20.0$ (RMSD of 16.390 and TM-score of 0.613).
This ablation study reveals that moderate regressor guidance ($\lambda=5.0$) yields the best results, optimizing enzyme activity while maintaining structural stability.

\section{Conclusion}
In this work, we propose $k_{cat}$Diffuser, a novel regressor-guided graph diffusion model designed to enhance enzyme turnover numbers while maintaining protein structural integrity. 
By reformulating enzyme mutation prediction as a protein inverse folding task, our approach establishes a direct link between structural prediction and functional optimization. 
Our evaluation demonstrates that $k_{cat}$Diffuser outperforms state-of-the-art methods across multiple metrics. 
While $k_{cat}$Diffuser exhibits higher computational complexity compared to some baseline models, this trade-off enables multi-site mutations and activity optimization capabilities not present in faster approaches. 
The model's moderate memory footprint also ensures practical deployability across various hardware configurations.
Thus, $k_{cat}$Diffuser represents an advancement in computational enzyme engineering, offering an efficient and targeted approach to enhancing enzyme activity. 
By enabling the prediction of beneficial multi-site mutations, our model addresses key challenges in enzyme optimization and opens new avenues for rational design of improved biocatalysts. 
Future work could focus on further optimizing the model's efficiency and exploring its applicability to a broader range of enzyme classes and reaction types.


\begin{thebibliography}{00}
\bibitem{b1}
K.~Chen and F.~H. Arnold, ``Engineering new catalytic activities in enzymes,'' \emph{Nature Catalysis}, vol.~3, no.~3, pp. 203--213, 2020.

\bibitem{b2}
P.~Wendering, M.~Arend, Z.~Razaghi-Moghadam, and Z.~Nikoloski, ``Data integration across conditions improves turnover number estimates and metabolic predictions,'' \emph{Nature Communications}, vol.~14, no.~1, p. 1485, 2023.

\bibitem{b3}
S.~Qiu, S.~Zhao, and A.~Yang, ``Dltkcat: deep learning-based prediction of temperature-dependent enzyme turnover rates,'' \emph{Briefings in Bioinformatics}, vol.~25, no.~1, p. bbad506, 2024.

\bibitem{b4}
F.~Li, L.~Yuan, H.~Lu, G.~Li, Y.~Chen, M.~K. Engqvist, E.~J. Kerkhoven, and J.~Nielsen, ``Deep learning-based k cat prediction enables improved enzyme-constrained model reconstruction,'' \emph{Nature Catalysis}, vol.~5, no.~8, pp. 662--672, 2022.

\bibitem{b5}
T.~Wang, G.~Xiang, S.~He, L.~Su, X.~Yan, and H.~Lu, ``Deepenzyme: a robust deep learning model for improved enzyme turnover number prediction by utilizing features of protein 3d structures,'' \emph{bioRxiv}, pp. 2023--12, 2023.

\bibitem{b6}
J.~Meier, R.~Rao, R.~Verkuil, J.~Liu, T.~Sercu, and A.~Rives, ``Language models enable zero-shot prediction of the effects of mutations on protein function,'' \emph{Advances in neural information processing systems}, vol.~34, pp. 29\,287--29\,303, 2021.

\bibitem{b7}
K.~Yi, B.~Zhou, Y.~Shen, P.~Li{\`o}, and Y.~Wang, ``Graph denoising diffusion for inverse protein folding,'' \emph{Advances in Neural Information Processing Systems}, vol.~36, 2024.

\bibitem{b8}
J.~Sohl-Dickstein, E.~Weiss, N.~Maheswaranathan, and S.~Ganguli, ``Deep unsupervised learning using nonequilibrium thermodynamics,'' in \emph{International conference on machine learning}.\hskip 1em plus 0.5em minus 0.4em\relax PMLR, 2015, pp. 2256--2265.

\bibitem{b9}
Y.~Song and S.~Ermon, ``Generative modeling by estimating gradients of the data distribution,'' \emph{Advances in neural information processing systems}, vol.~32, 2019.

\bibitem{b10}
P.~Dhariwal and A.~Nichol, ``Diffusion models beat gans on image synthesis,'' \emph{Advances in neural information processing systems}, vol.~34, pp. 8780--8794, 2021.

\bibitem{b11}
K.~Black, M.~Janner, Y.~Du, I.~Kostrikov, and S.~Levine, ``Training diffusion models with reinforcement learning,'' \emph{arXiv preprint arXiv:2305.13301}, 2023.

\bibitem{b12}
H.~He, C.~Bai, K.~Xu, Z.~Yang, W.~Zhang, D.~Wang, B.~Zhao, and X.~Li, ``Diffusion model is an effective planner and data synthesizer for multi-task reinforcement learning,'' \emph{Advances in neural information processing systems}, vol.~36, 2024.

\bibitem{b13}
J.~S. Lee, J.~Kim, and P.~M. Kim, ``Score-based generative modeling for de novo protein design,'' \emph{Nature Computational Science}, vol.~3, no.~5, pp. 382--392, 2023.

\bibitem{b14}
E.~Hoogeboom, V.~G. Satorras, C.~Vignac, and M.~Welling, ``Equivariant diffusion for molecule generation in 3d,'' in \emph{International conference on machine learning}.\hskip 1em plus 0.5em minus 0.4em\relax PMLR, 2022, pp. 8867--8887.

\bibitem{b15}
K.~E. Wu, K.~K. Yang, R.~van~den Berg, S.~Alamdari, J.~Y. Zou, A.~X. Lu, and A.~P. Amini, ``Protein structure generation via folding diffusion,'' \emph{Nature communications}, vol.~15, no.~1, p. 1059, 2024.

\bibitem{b16}
J.~Ingraham, V.~Garg, R.~Barzilay, and T.~Jaakkola, ``Generative models for graph-based protein design,'' \emph{Advances in neural information processing systems}, vol.~32, 2019.

\bibitem{b17}
B.~Jing, S.~Eismann, P.~Suriana, R.~J.~L. Townshend, and R.~Dror, ``Learning from protein structure with geometric vector perceptrons,'' in \emph{International Conference on Learning Representations}, 2020.

\bibitem{b18}
C.~Tan, Z.~Gao, J.~Xia, B.~Hu, and S.~Z. Li, ``Generative de novo protein design with global context,'' \emph{arXiv preprint arXiv:2204.10673}, 2022.

\bibitem{b19}
A.~Strokach, D.~Becerra, C.~Corbi-Verge, A.~Perez-Riba, and P.~M. Kim, ``Fast and flexible protein design using deep graph neural networks,'' \emph{Cell systems}, vol.~11, no.~4, pp. 402--411, 2020.

\bibitem{b20}
J.~Dauparas, I.~Anishchenko, N.~Bennett, H.~Bai, R.~J. Ragotte, L.~F. Milles, B.~I. Wicky, A.~Courbet, R.~J. de~Haas, N.~Bethel \emph{et~al.}, ``Robust deep learning--based protein sequence design using proteinmpnn,'' \emph{Science}, vol. 378, no. 6615, pp. 49--56, 2022.

\bibitem{b21}
Z.~Gao, C.~Tan, P.~Chac{\'o}n, and S.~Z. Li, ``Pifold: Toward effective and efficient protein inverse folding,'' \emph{arXiv preprint arXiv:2209.12643}, 2022.

\bibitem{b22}
D.~Heckmann, C.~J. Lloyd, N.~Mih, Y.~Ha, D.~C. Zielinski, Z.~B. Haiman, A.~A. Desouki, M.~J. Lercher, and B.~O. Palsson, ``Machine learning applied to enzyme turnover numbers reveals protein structural correlates and improves metabolic models,'' \emph{Nature communications}, vol.~9, no.~1, p. 5252, 2018.

\bibitem{b23}
A.~Kroll, Y.~Rousset, X.-P. Hu, N.~A. Liebrand, and M.~J. Lercher, ``Turnover number predictions for kinetically uncharacterized enzymes using machine and deep learning,'' \emph{Nature communications}, vol.~14, no.~1, p. 4139, 2023.

\bibitem{b24}
C.~A. Orengo, A.~D. Michie, S.~Jones, D.~T. Jones, M.~B. Swindells, and J.~M. Thornton, ``Cath--a hierarchic classification of protein domain structures,'' \emph{Structure}, vol.~5, no.~8, pp. 1093--1109, 1997.

\bibitem{b25}
H.~M. Berman, J.~Westbrook, Z.~Feng, G.~Gilliland, T.~N. Bhat, H.~Weissig, I.~N. Shindyalov, and P.~E. Bourne, ``The protein data bank,'' \emph{Nucleic acids research}, vol.~28, no.~1, pp. 235--242, 2000.

\bibitem{b26}
A.~Chang, L.~Jeske, S.~Ulbrich, J.~Hofmann, J.~Koblitz, I.~Schomburg, M.~Neumann-Schaal, D.~Jahn, and D.~Schomburg, ``Brenda, the elixir core data resource in 2021: new developments and updates,'' \emph{Nucleic acids research}, vol.~49, no.~D1, pp. D498--D508, 2021.

\bibitem{b27}
B.~Hie, S.~Candido, Z.~Lin, O.~Kabeli, R.~Rao, N.~Smetanin, T.~Sercu, and A.~Rives, ``A high-level programming language for generative protein design,'' \emph{bioRxiv}, pp. 2022--12, 2022.

\bibitem{b28}
J.~Jumper, R.~Evans, A.~Pritzel, T.~Green, M.~Figurnov, O.~Ronneberger, K.~Tunyasuvunakool, R.~Bates, A.~{\v{Z}}{\'\i}dek, A.~Potapenko \emph{et~al.}, ``Highly accurate protein structure prediction with alphafold,'' \emph{nature}, vol. 596, no. 7873, pp. 583--589, 2021.

\bibitem{b29}
W.~Kabsch, ``A solution for the best rotation to relate two sets of vectors,'' \emph{Acta Crystallographica Section A: Crystal Physics, Diffraction, Theoretical and General Crystallography}, vol.~32, no.~5, pp. 922--923, 1976.

\bibitem{b30}
Y.~Zhang and J.~Skolnick, ``Scoring function for automated assessment of protein structure template quality,'' \emph{Proteins: Structure, Function, and Bioinformatics}, vol.~57, no.~4, pp. 702--710, 2004.

\bibitem{b31}
J.~Xu and Y.~Zhang, ``How significant is a protein structure similarity with tm-score= 0.5?'' \emph{Bioinformatics}, vol.~26, no.~7, pp. 889--895, 2010.

\bibitem{b32}
A.~Vaswani, N.~Shazeer, N.~Parmar, J.~Uszkoreit, L.~Jones, A.~N. Gomez, {\L}.~Kaiser, and I.~Polosukhin, ``Attention is all you need,'' \emph{Advances in neural information processing systems}, vol.~30, 2017.

\bibitem{b33}
T.~N. Kipf and M.~Welling, ``Semi-supervised classification with graph convolutional networks,'' \emph{arXiv preprint arXiv:1609.02907}, 2016.

\bibitem{b34}
J.~Song, C.~Meng, and S.~Ermon, ``Denoising diffusion implicit models,'' \emph{arXiv preprint arXiv:2010.02502}, 2020.


\end{thebibliography}
\end{document}